\documentclass[a4paper,11pt]{article}
\usepackage{pos}
\usepackage{amsmath}

\DeclareMathOperator\Tr{Tr}
\usepackage[numbers]{natbib}

\title{Tensor Renormalization Group Methods for Quantum Real-time Evolution}

\author*[a]{Michael Hite}
\author[a]{Yannick Meurice}

\affiliation[a]{University of Iowa,\\
  30 North Dubuque St., Iowa City, Iowa 52242}

\emailAdd{michael-hite@uiowa.edu}
\emailAdd{yannick-meurice@uiowa.edu}

\abstract{Ab-initio calculations of real-time evolution for lattice gauge theory have very interesting potential applications but present challenging computational aspects. We show that tensor renormalization group methods developed in the context of  Euclidean-time lattice field theory can be applied to calculation of Trotterized evolution operators at real-time. We discuss the optimization of truncation procedures for the real-time equivalent of the partition function. We apply the numerical methods to the 1D Quantum Ising Model with an external transverse field in the ordered phase and compare with universal quantum computing for $N_{s}=4$ and 8 sites.}

\FullConference{The 40th International Symposium on Lattice Field Theory (Lattice 2023)\\
July 31st - August 4th, 2023\\
Fermi National Accelerator Laboratory\\}

\begin{document}
\maketitle

\section{Introduction}
Lattice gauge theory simulations are well known to be computationally intensive. Traditionally, these problems have been approximated in Euclidean-time via course-graining methods such as the Tensor Renormalization group (TRG) \cite{TensorReview}. Though promising, TRG methods have drawbacks such as microscopic nonphysical behavior for larger systems \cite{TNR}. Recently, with quantum computers, we have the potential to simulate real-time scattering processes, but, in the noisy-intermediate quantuam (NISQ) era, noise and decoherence limits us to small systems. Thus, the path forward is not well defined. For some models, such as the 1D Quantum Ising Model (QIM), there exist an exact mapping to a classical model, which means we can treat the quantum model exactly as we would a classical model. So classical methods, such as TRG, can be applied to them in real-time. Using the 1D QIM as our test bed, we will lay some rough framework for quantum real-time evolution using the TRG method known as Higher-Order TRG (HOTRG) \cite{HOTRG}.

The paper is organized as follows: In section 2 we introduce the 1D QIM and 2D Classical Ising Model (CIM) with the mapping of the classical transfer matrix to the quantum real-time evolution operator. We then re-express the transfer matrix in terms of tensors to construct the equivalent tensor network for a given number of sites to do HOTRG. In section 3 we compare the time evolution operator of HOTRG with exact diagonalization for increasing approximation, as well as compare the algorithm's results for real and imaginary time evolution. Finally, in section 4 we conclude and propose future work.

\section{Theory}
\subsection{1D Quantum Ising Model to 2D Classical Ising Model}
Consider a ring of $N_s$ spin-1/2 particles, where interactions occur among nearest-neighbors and a transverse magnetic field. The Hamiltonian for this quantum system in the Pauli-$x$ basis is given by
\begin{equation}
    \hat{H}_\text{QIM} = - \sum_{i=1}^{N_s} \left( \lambda\hat{\sigma}^x_{i+1}\hat{\sigma}^x_i + \hat{\sigma}^z_i \right).
\end{equation}
where $\lambda$ controls the strength of nearest-neighbor interactions, the coupling to the transverse magnetic field is set to one, and $\hat{\sigma}_{N_s + 1} = \hat{\sigma}_1$. On a quantum computer, we are restricted to applying non-commuting operators separately. So, to evolve a small time interval $\Delta t$, the Trotter time evolution operator (with first order correction) is
\begin{equation}
    U(\Delta t) = e^{-i\Delta t\hat{H}_\text{T}/2}e^{-i\Delta t\hat{H}_\text{NN}}e^{-i\Delta t\hat{H}_\text{T}/2},
    \label{TrotTEO}
\end{equation}
where $\hat{H}_\text{NN}$ and $\hat{H}_\text{T}$ are the nearest-neighbor and transverse field terms of the Hamiltonian respectively.

The 1D QIM can be mapped to the 2D anisotropic classical Ising model. For an $N_\tau \times N_s$-lattice with a spin variable $\sigma = \pm 1$ attached to each site and only nearest neighbor interactions, the Hamiltonian for the system is given by
\begin{equation}
    H_\text{CIM} = -\sum_{i=1}^{N_s}\sum_{j=1}^{N_\tau} \left( J_s \sigma_{i,j} \sigma_{i+1,j} + J_\tau \sigma_{i,j}\sigma_{i,j+1}\right),
\end{equation}
where $J_s(J_\tau)$ controls the strength of the horizontal (vertical) interactions. The partition function for the system is simply
\begin{equation}
    Z = \sum_{\{\sigma\}} e^{-\beta H_\text{CIM}} = \text{Tr }\left( \hat{\mathbb{V}}^{N_\tau}\right),
\end{equation}
where $\hat{\mathbb{V}}$ is a transfer matrix that lives on a single horizontal slice. The isotropic case was solved exactly by Lars Onsager \cite{Onsager} and Bruria Kaufman \cite{Kaufman} in the 1940's. It can be shown that the transfer matrix is
\begin{equation}
    \hat{\mathbb{V}} = \hat{\mathbb{V}}_2^{1/2} \hat{\mathbb{V}}_1 \hat{\mathbb{V}}_2^{1/2}
\end{equation}
where
\begin{equation}
    \begin{split}
        &\hat{\mathbb{V}}_2 = \left(\sinh2\beta_\tau)\right)^{N_s/2} e^{\beta_\tau^* \sum_{i=1}^{N_s} \hat{\sigma}^z_i}, \\
        &\hat{\mathbb{V}}_1 = e^{\beta_s \sum_{i=1}^{N_s} \hat{\sigma}^x_{i+1}\hat{\sigma}^x_{i}},
    \end{split}
\end{equation}
$\beta_s=\beta J_s,\; \beta_\tau = \beta J_\tau$, and $\tanh\beta_\tau = e^{-2\beta_\tau^*}$ \cite{TensorReview}. Under the mapping $\beta_\tau^* \rightarrow i\Delta t$ and $\beta_s \rightarrow i\lambda \Delta t$, while also ignoring the prefactor in $\hat{\mathbb{V}}_2$, we recover the Trotter time evolution operator in Equation \ref{TrotTEO} from the transfer matrix. Thus we have a connection between the quantum time evolution operator and the classical transfer matrix from statistical mechanics.

\subsection{The Higher-Order Tensor Renormalization Group}
\begin{figure}
    \centering
    \includegraphics[scale=0.5]{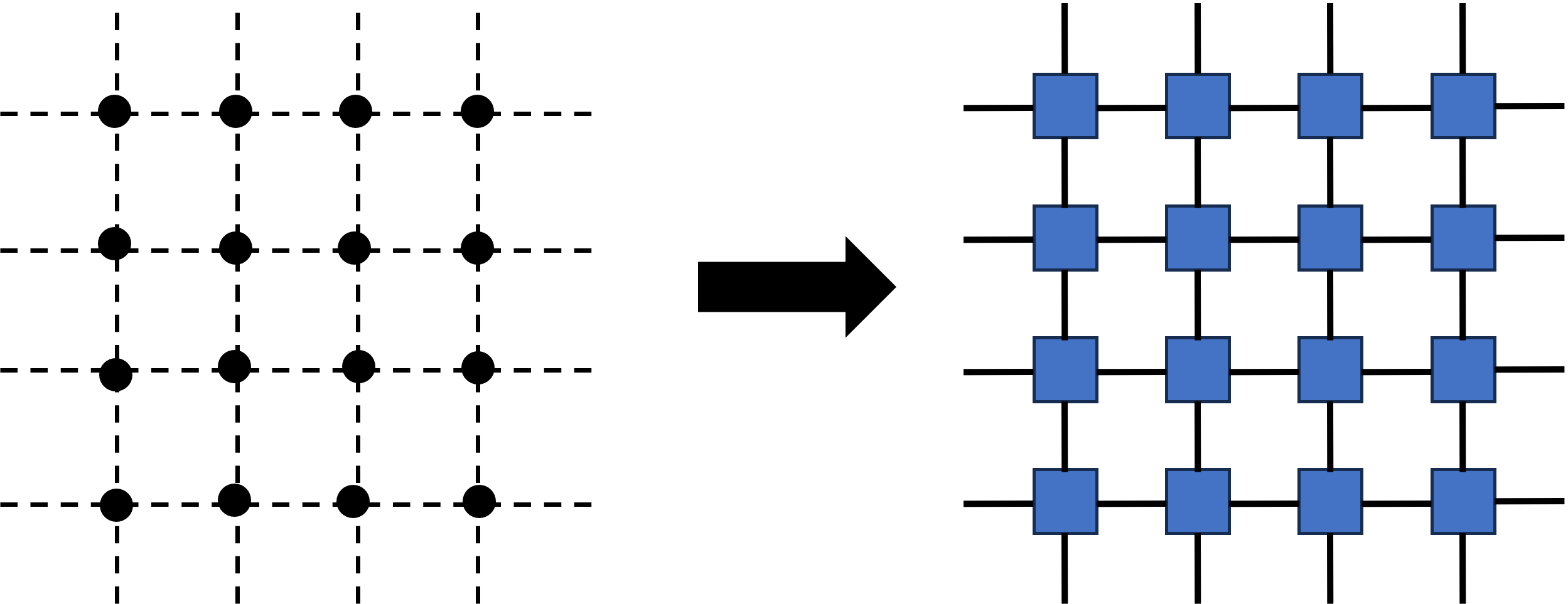}
    \caption{A $4\times4$ CIM represented as a tensor network.}
    \label{fig:lat-to-tn}
\end{figure}

In statistical mechanics, there are many ways to represent the partition function. We have shown the transfer matrix representation. Yet, given the nature of the problem, it can be useful to represent the partition function as the trace of a product of tensors. For the 2D anisotropic CIM, the rank-4 tensor
\begin{equation}
T_{ijkl}^{(0)} \equiv\left(\sqrt{\tanh(\beta_{s})}\right)^{i+j}\left(\sqrt{\tanh(\beta_{\tau})}\right)^{k+l}\delta\left[\left(i+j+k+l\right)\%2\right],
\label{eq:T-tensor}
\end{equation}
is attached to a site, and the indices set on the left and right (spatial), and bottom and top (temporal) legs respectively (Blue tensor in Figure \ref{fig:lat-to-tn}).

\begin{figure}
    \centering
    \includegraphics[scale=0.6]{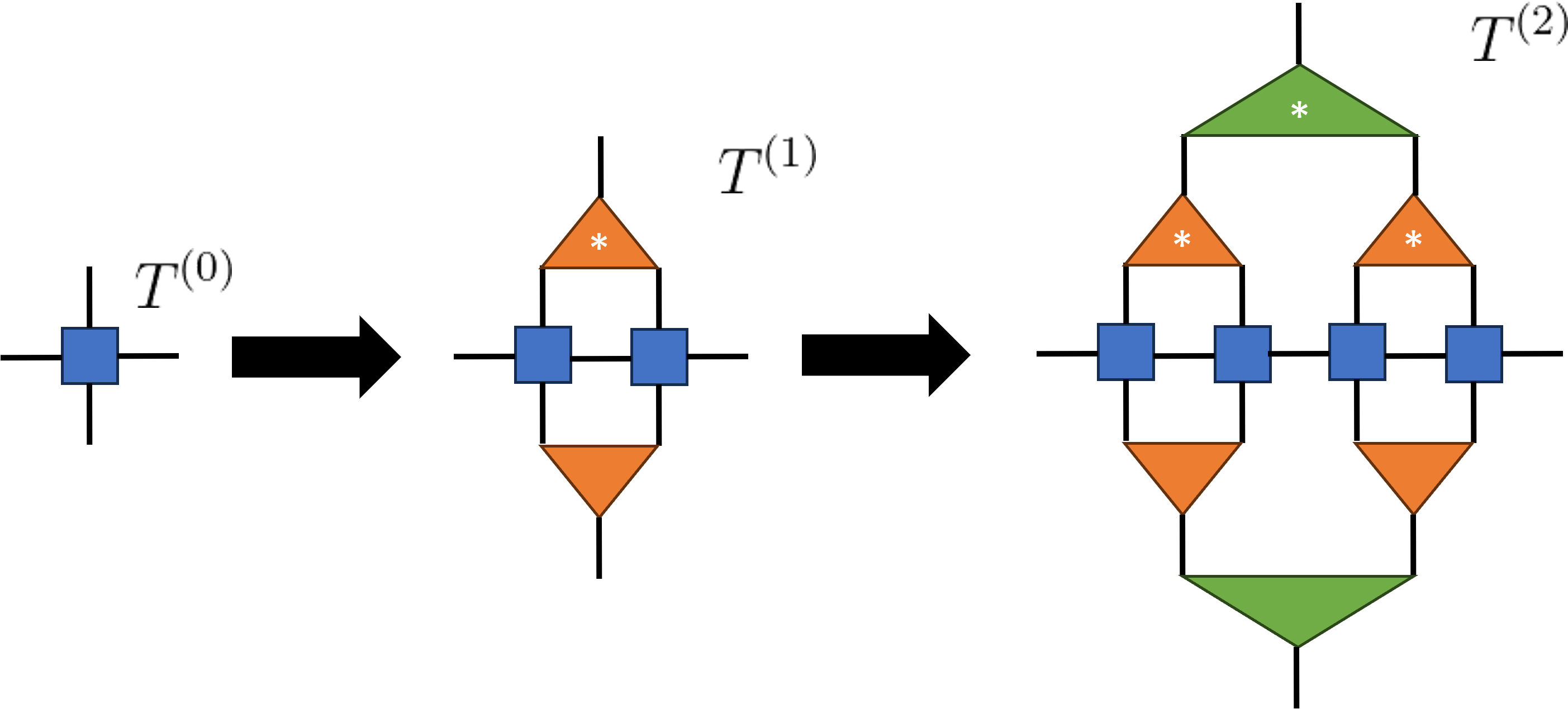}
    \caption{Iterations 0, 1, and 2 of the HOTRG algorithm. Dimensionality goes like $(2,2,2,2), (2,2,4,4),$ and $(2,2,16,16)$. The orange tensor corresponds to $U^{(1)}$, and the green tensor the $U^{(2)}$. The transfer matrix is obtained by tracing over the spatial legs. The star corresponds to the complex conjugate of $U^{(n)}$.}
    \label{fig:HOTRG-alg}
\end{figure}

The HOTRG algorithm works as follows. Beginning with a single $T^{(0)}$, we seek to construct a tensor $U^{(1)}$ (the orange tensor in Figure \ref{fig:HOTRG-alg}) so that the product $T^{(0)}T^{(0)}$ will be itself a rank-4 tensor. $U^{(1)}$ is a tensor of eigenvectors of a matrix $Q^{(0)}$ whose trace is equivalently the norm of $T^{(0)}T^{(0)}$, and contains all the information to tie together the two bottom (or top) legs. Different choices of $Q^{(0)}$ can be used depending upon the context of the problem \cite{KaufmansSolution}, which we will explore later on. The dimensions of $U^{(1)}$ are $(2, 2, 4)$, making the dimensions of the new tensor $T^{(1)}$ $(2,2,4,4)$. Thus, we now have a single rank-4 tensor for a 2-site model. Repeating the procedure with $T^{(1)}$, $U^{(2)}$ will have dimensions $(4,4,16)$, and the new tensor $T^{(2)}$ will have dimensions $(2,2,16,16)$ that represents a 4-site model. The algorithm is iterated $n = \log_2(N_s)$ times to produce an exact rank-4 tensor for an $N_s$-site model.

We can approximate the network by only keeping the largest contributing eigenstates in the construction of $U^{(n)}$. For instance, if we set a cutoff $d_\text{bound} = 14$, then $U^{(2)}$ will have dimensions $(4,4,14)$, and $T^{(2)}$ will have dimensionality $(2,2,14,14)$. This $d_\text{bond}$ will have to be fine tuned to get in order to get a sufficient approximation for a given lattice size. Finally, making the appropriate substitutions $\beta_\tau \rightarrow \beta_\tau - i\pi/4$ and $\beta_s \rightarrow i\lambda t$, we get the time evolution operator by tracing over the spatial legs and multiplying by an overall factor of $(e^{i\Delta t}\cosh(i\lambda \Delta t))^{N_s}$.

\section{Results}
\begin{figure}[h!]
    \centering
    \includegraphics[scale=0.5]{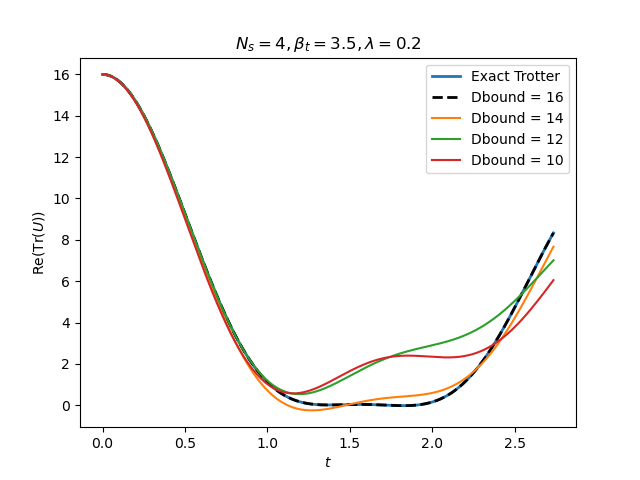}\includegraphics[scale=0.5]{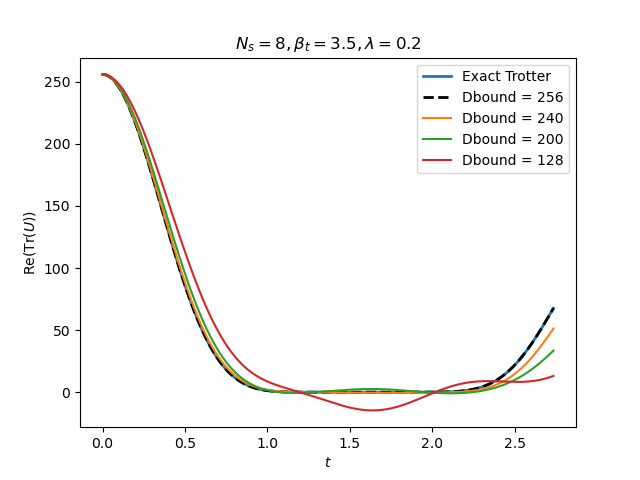}
    \includegraphics[scale=0.5]{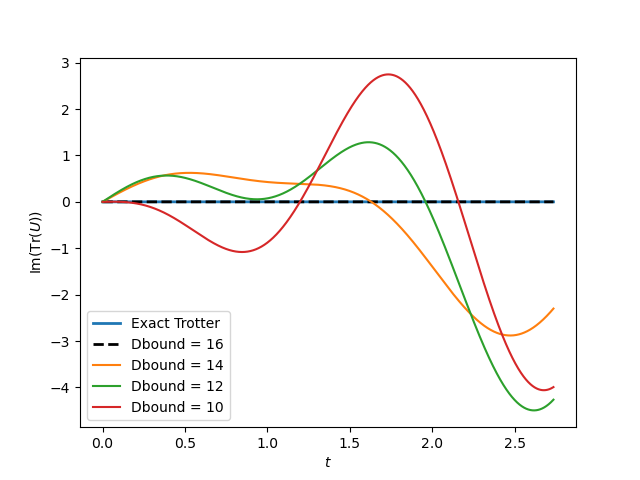}\includegraphics[scale=0.5]{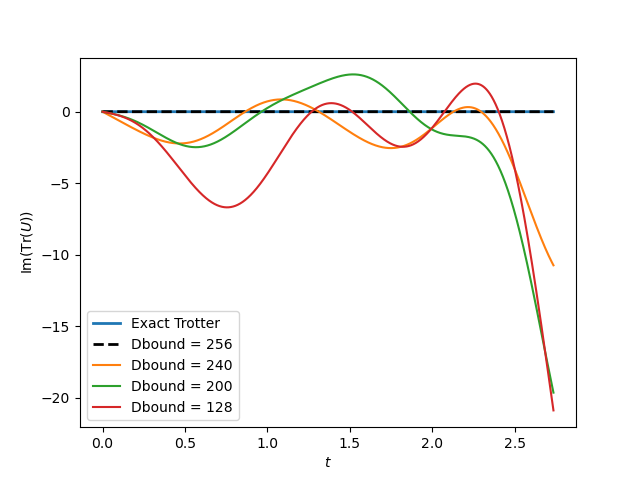}
    \caption{Real-time evolution of the real and imaginary part of $2^{N_s} \Tr($U$) / D_\text{bound}$ for $N_s = 4$ (column 1) and 8 sites (column 2). The blue line corresponds to exact diagonalization. The dashed line is $D_\text{bound} = 16$ and 256, corresponding to no cutoff (exact HOTRG). The other lines introduce a cutoff.}
    \label{fig:TrU}
\end{figure}

Now that we have the HOTRG algorithm developed in the context of real-time, we can perform some preliminary tests. A simple quantity to measure is the trace of the time evolution operator $U$, as it is the real-time equivalent of the classical partition function. Figure \ref{fig:TrU} shows the time evolution of Tr($U$) for $\lambda = 0.2$ and $\beta_t = 3.5$ ($\Delta t = e^{-2\beta_t} \approx 0.0009$) for 4 and 8 sites. We included a factor of $2^{N_s}/d_\text{bond}$ so that Tr($U^0$) = 16 in the truncated case.  We see that for $d_\text{bound} = 16$ and 256, which is exact HOTRG (no cutoff), agrees with exact diagonalization. Once we truncate, we can see the effect on Tr($U$), especially the imaginary part.

\begin{figure}[h!]
    \centering
    \includegraphics[scale=0.5]{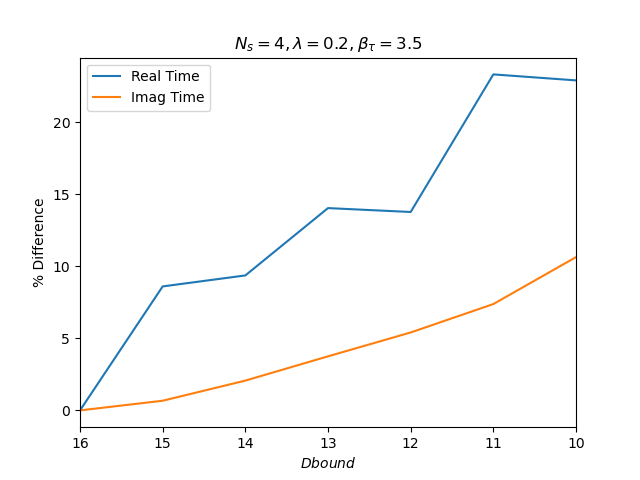}\includegraphics[scale=0.5]{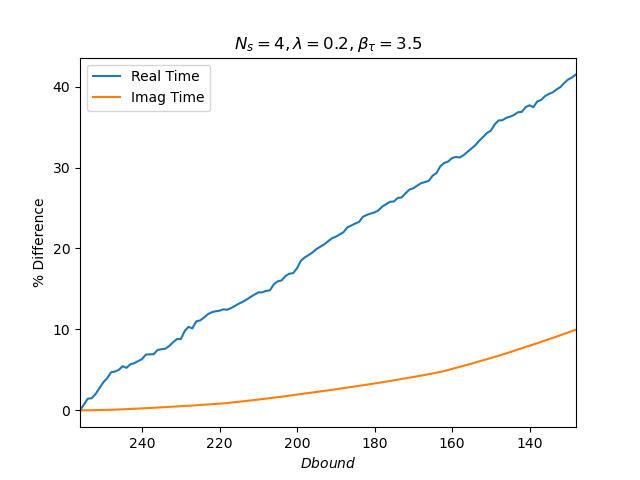}
    \caption{Comparison of percent difference of Tr$(U(\Delta t)^N)$ for $\Delta t = e^{-7}$ and $N = 500$ time steps ($t\approx 0.46$). The plots are for $N_s =4$ and $8$ sites respectively. We see that HOTRG is far more stable in the context of Euclidean time than real time.}
    \label{fig:rte-ite-comparison}
\end{figure}

\begin{figure}[h!]
    \centering
    \includegraphics[scale=0.5]{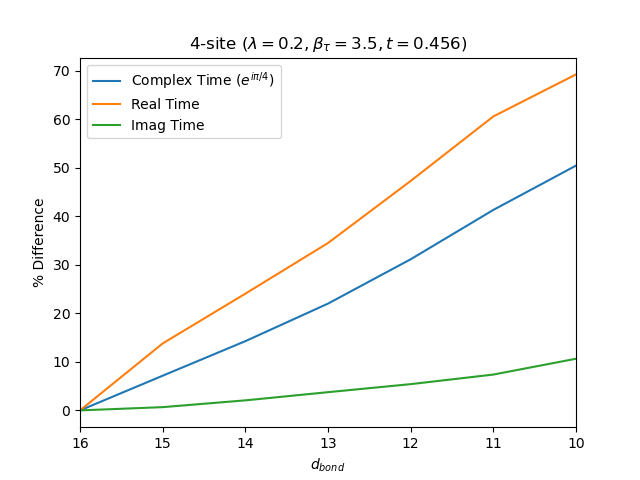}\includegraphics[scale=0.5]{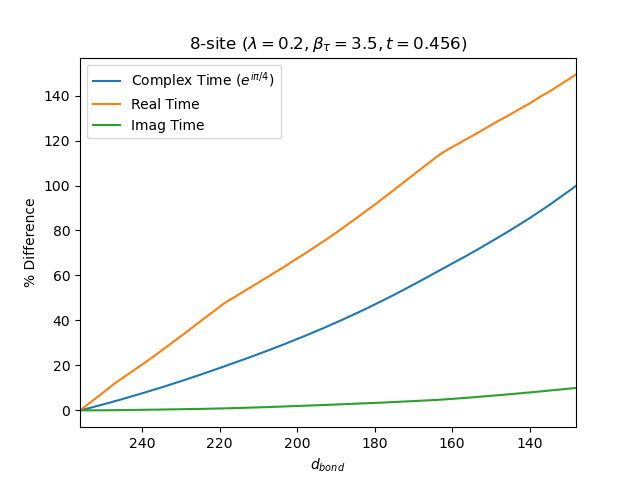}
    \caption{For a new choice of $Q^{(0)}$ that accounts for complex elements of $T^{(0)}$, we look at percent difference at 500 times steps ($t \approx 0.46$) for real, complex and Euclidean time.}
    \label{fig:new-Q}
\end{figure}

We can also look at the difference from the exact result (real and imaginary) in the cases of real and Euclidean-time evolution. For the Euclidean-time case, we neglect the factor of $2^{N_s}/d_\text{bond}$. In Figure \ref{fig:TrU} there is noticeable deviation when we approach 500 times steps ($t \approx 0.46$), so we can compare the deviation for real and Euclidean time at that point (Figure \ref{fig:rte-ite-comparison}). We see that in real-time the HOTRG result is more sensitive to the removal of states than Euclidean-time. Figure \ref{fig:rte-ite-comparison} also confirms the validity in the use of HOTRG in Euclidean time, as the percent difference remains well below 10\% in the 4-site case even when half of the states are removed. We can choose a different $Q^{(0)}$ developed specifically for tensors with complex elements \cite{KaufmansSolution}. Figure \ref{fig:new-Q} shows results for real and Euclidean time for the different $Q^{(0)}$ as well as for a complex time ($\Delta \tau \rightarrow \Delta t e^{i\pi/4)}$). The percent error jumps dramatically for the real-time case, which we hope to investigate further.

\section{Conclusion}
We have shown very early work in using TRG methods for quantum real-time evolution. Compared to other TRG methods, HOTRG allows for coarse graining in a single dimension, improving approximation schemes. For $N_s = 4$ and 8 sites, HOTRG agrees with exact diagonalization when keeping all states, and shows expected behavior when truncating. As HOTRG was created in the context of Euclidean-time, we see that it performs far better in Euclidean-time than in real-time. We then used a different $Q^{(0)}$ metric defined specifically to handle complex entries in $T^{(0)}$, but only saw minimal gains for given values of $d_\text{bond}$. We see the algorithm still performs fairly well when we rotate by $\pi/4$, but struggles once we drop onto the imaginary-axis (real-time).

In future work we seek to answer the following questions: What happens to the HOTRG algorithm as we rotate from imaginary-time to real-time? In real-time, does the HOTRG algorithm select the low energy/momentum states? If so, then what is the connection to the Wilsonian Renormalization Group \cite{WILSON1974}? If not, is there a different metric $Q^{(0)}$ that could be used to do so? Can the same projection used in statistical mechanics be used for real-time evolution? Finally, provided translation invariance, can we use HOTRG to simulate quantum wave packets centered at low momenta (scattering) and compare to results of quantum simulation \cite{QSimScatt}?

\section*{Acknowledgements}
This work is supported in part by the Department of Energy under Award Numbers DE-SC0019139 and DE-SC0010113. Michael Hite was additionally supported in part by NSF award DMR-1747426.

\bibliographystyle{unsrt} 
\bibliography{refs} 

\end{document}